\documentclass[12pt]{iopart}

\usepackage{iopams}  

\usepackage{graphicx}
\usepackage{tensor}

\newcommand{\bra}[1]{\left\langle{#1}\right\vert}
\newcommand{\ket}[1]{\left\vert{#1}\right\rangle}
\newcommand{\ketbra}[2]{|#1\rangle \langle#2|}
\newcommand{\openone}{1 \hspace{-1.0mm}  {\bf l}}

\begin{document}

\title[]{Experimental characterization of photonic fusion using fiber sources}

\author{B. Bell,$^1$ A. Clark,$^1$ M. S. Tame,$^2$ M. Halder,$^1$ J. Fulconis,$^1$ \\ W. Wadsworth,$^3$ and J. Rarity$^1$}

\address{$^1$Centre for Communications Research, Department of Electrical and Electronic Engineering, University of Bristol, Merchant Venturers Building, Woodland Road, Bristol, BS8 1UB, UK \\ $^2$Quantum Optics and Laser Science, The Blackett Laboratory, Imperial College London, Prince Consort Road, SW7 2BW, United Kingdom \\ $^3$Centre for Photonics and Photonic Materials, Department of Physics, University of Bath, Claverton Down, Bath, BA2 7AY, United Kingdom}
\begin{abstract}
We report the fusion of photons from two independent photonic crystal fiber sources into polarization entangled states using a fiber-based polarizing beam splitter. We achieve fidelities of up to $F~=~0.74~\pm~0.01$ with respect to the maximally entangled Bell state $\ket{\phi^+}$ using a low pump power of $5.3$mW with a success rate of $3.2$ four-fold detections per second. By increasing the pump power we find that success rates of up to 111.6 four-folds per second can be achieved, with entanglement still present in the fused state. We characterize the fusion operation by providing a full quantum process reconstruction. Here a model is developed to describe the generation of entanglement, including the main causes of imperfection, and we show that this model fits well with the experimental results. Our work shows how non-ideal settings limit the success of the fusion, providing useful information about the practical requirements for an operation that may be used to build large entangled states in bulk and on-chip quantum photonic waveguides.
\end{abstract}

\pacs{42.50.-p, 42.70.Qs, 42.81.-i}
\maketitle

\section{Introduction}
The controlled generation of quantum entanglement is an essential process for performing a wide range of tasks in the field of quantum information~\cite{NC}. Quantum communication protocols such as teleportation~\cite{Bennett}, key distribution~\cite{Ekert} and dense~coding~\cite{Dense} all require the generation of entangled states. Entanglement is also generated during quantum computation~\cite{QCent} and in the simulation of many-body quantum systems~\cite{MBQS}. So far, small-sized entangled states have been generated in a wide range of physical setups, with photonic systems representing one of the most promising due to their speed and flexibility. Recently much attention has been focused on generating multiqubit entangled states such as cluster~\cite{cluster,PANPRA}, graph~\cite{graph} and Dicke~\cite{dicke} states, as well as those with more unusual structures and correlations~\cite{rad} in a probabilistic fashion. Schemes that allow for the deterministic generation of smaller two-qubit entangled states have also been realized~\cite{panp,Barz2,Hubel}, although low generation rates make these approaches challenging at present for building up to larger multiqubit entangled states. Photonic setups that generate states probabilistically, although inherently non-scalable, provide a readily available test-bed for probing the unique properties of quantum systems. Current experiments are however limited to entangled states of ten-qubits or less~\cite{panp2}. Improving photon generation rates using new types of sources and understanding better the practical requirements for generating high-quality entanglement between these sources may open up access to even larger entangled states with more complex structures. This would enable the testing of quantum protocols and probing physical phenomena that only more sizable quantum systems are able to support. 

In this work we report the first experiment to fuse photons from two independent photonic crystal fiber (PCF) sources into polarization entangled states. The fusion operation we demonstrate could be used to generate larger multiqubit entangled states, such as cluster and graph states for use in future quantum photonic technologies~\cite{OBRIEN3}, including quantum communication and computation. We also introduce a novel method to characterize the fusion process that could be applied to a variety of different setups, including bulk, waveguide and on-chip photonic systems. While the fusion process is probabilistic in our experiment, we comment on how it may be made deterministic in the long term for the purposes of building large scalable entangled quantum systems. Our PCF sources have a very high photon generation rate compared to commonly used nonlinear crystal sources~\cite{Hald}. Moreover, they make use of a zero-slope section of the phase-matching curve to produce photons in an intrinsically pure state, enabling good quality quantum interference without the need for spectral filtering (with inherent loss) resulting in high count rates~\cite{Hald,Clark}. We detect~90,000 coincidences per second from photon pairs produced with a pump laser power of 10.5mW. In this context, we investigate the fusion of photons from two independent sources in detail and provide a full quantum process reconstruction. We do this by developing a theoretical model to describe the entangling operation that includes the main causes of imperfection between the photons being fused. We find that our model fits well with the experimental results. Our work shows how non-ideal settings limit the success of the fusion operation and provides detailed information about the practical requirements for using it to generate large multiqubit entangled states in bulk~\cite{Hald,Clark} and on-chip quantum photonic settings~\cite{OBRIEN2,Sansoni,DJBV,PBS,Mat}. The fusion is performed with a fiber polarizing beamsplitter (FPBS), which is better suited than the bulk optics equivalent for integration into more complex schemes and scaling up to generate larger entangled states, where space constraints and coupling stability become important considerations. As it is waveguide-based, it is also interferometrically stable and well suited to the generation of resource states that are to be transmitted and received over fiber networks, for use in distributed quantum communication and networking protocols.

\section{Experimental setup}

\begin{figure*}[t]
\begin{minipage}{14cm}
\hspace*{2cm}\includegraphics[width=13cm]{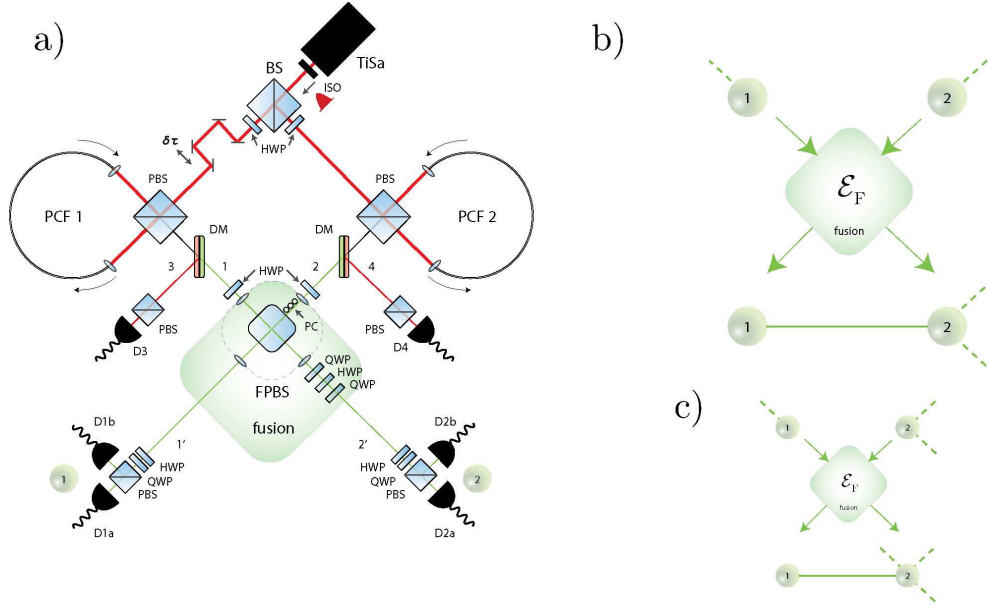}
\end{minipage}
\caption{{\bf (a):}~Experimental setup {\bf (b):}~Fusion operation used on two linear cluster states to make a new linear cluster state with an attached qubit (qubit 1) that can be used for subsequent fusions after a Hadamard operation is applied or for decoherence protection~\cite{DFS}. {\bf (c):}~Fusion operation used on two linear cluster states to make a new two-dimensional cluster state.}
\label{setup}
\end{figure*}

Our setup for fusing photons from two independent PCF sources is depicted in Fig.~\ref{setup}. A Ti:sapphire laser emits 8 nm pulses at a wavelength of 724 nm with a repetition rate of 80 MHz, which are then filtered to 1 nm. The pulses are split at a 50:50 beamsplitter (BS) and rotated to horizontal polarization by halfwave plates (HWPs) in both arms, then passed through polarizing beamsplitters (PBSs) and launched into the two PCFs. The fibers each produce a pair of photons (signal and idler) polarized orthogonally to the pump via four-wave mixing with nondegenerate wavelengths of 625 nm for the signals and 860 nm for the idlers. While these wavelengths can be tuned by changing the wavelength of the pump pulse, the photons are only emitted in an intrinsically pure state at this section of the phase-matching curve~\cite{Hald,Clark}. A $90^{\circ}$ twist in the fiber ensures these photons exit the PCFs horizontally polarized, with an aspheric lens directing them into the polarizing beamplitters (PBSs), which transmit them. The PBS helps to filter out the pump pulses, which exit the fiber vertically polarized, and unpolarized background such as Raman scattering. An isolator (ISO) at the output of the laser, which allows only one-way transmission, blocks any reflected pulses from entering back into the laser. The photon pairs are then separated into different paths using dichroic mirrors (DM).  The separated idler photons pass through a long- and short-pass filter tilted to give a tunable transmission window of~4 nm FWHM, which transmits the idler whilst removing as much Raman background as possible. They are then collected in single mode fibers followed by additional PBSs which remove background from reflected light in the PCFs. Multi-mode fibers are used to couple the idler to the detectors, where detections at D3 and D4 are used to herald the generation of the signal photons. The signal photons are rotated to diagonal polarization by HWPs and pass through a 40 nm FWHM bandpass filter. This bandwidth is large compared to the signal photon because the intrinsically pure state phase-matching makes narrow filtering unnecessary, and the filter only needs to remove any remaining light from the bright pump beam. They are then collected into single-mode fibers to guarantee optimal spatial overlap on the FPBS. Fiber polarization controllers (PCs) compensate for the effects on the polarization of strain-induced birefringence in the fibers. A quarter wave plate (QWP), HWP, QWP chain on mode 2' compensates any unwanted phase from the FPBS. Multi-mode fibers are used to couple the photons to the detectors. The detection of a photon at detector D1a (or D1b) with detector D2a (or D2b) together with detections at detectors D3 and D4 are used in a fourfold coincidence circuit to register a successful fusion. Here, the use of two detectors on each signal mode enables active filtering of higher-order photon emissions from four-wave mixing in the PCFs. 

\section{Fusion operation}
The fiber polarizing beamsplitter (FPBS) depicted in Fig.~\ref{setup} transmits horizontal polarized photons (H) and reflects vertical polarized photons (V). Ideally it can be described by a unitary operation with respect to the reduced two-photon basis ${\cal T}_p:=\{\ket{HH},\ket{HV},\ket{VH}, \ket{VV}, \ket{(HV)0}, \ket{0(HV)} \}$, where $\ket{(HV)0}$ ($\ket{0(HV)}$) represents two photons of different polarization in the first (second) mode and the vacuum, $\ket{0} \equiv \ket{vac}$, in the second (first). In the case that we `monitor' the number of photons present in the output ports,~{\it i.e.}~we gain information about the photon number but not the polarization, then the FPBS is described by the non-unitary trace-preserving channel ${\cal E} (\rho)=\sum_{i=0}^1 \hat{E}_i \rho \hat{E}_i^\dag$, with $\sum_{i=0}^1\hat{E}_i^\dag \hat{E}_i=\openone$. Here the Kraus operator $\hat{E}_0=\ketbra{HH}{HH}+\ketbra{VV}{VV}+\ketbra{HV}{0(HV)}+\ketbra{VH}{(HV)0}$ describes an operation into the `coincidence basis', or subspace of the Hilbert space with states having a single photon in each output mode ${\cal C}_b:=\{\ket{HH},\ket{HV},\ket{VH}, \ket{VV} \}$, where ${\cal C}_b \subset {\cal T}_p$. The second Kraus operator $\hat{E}_1=\ketbra{0(HV)}{HV}+\ketbra{(HV)0}{VH}$ describes an operation out of the subspace defined by the coincidence basis ${\cal C}_b$ and into a subspace of the basis ${\cal T}_p$ with states having two photons in one of the output modes. As all operations are limited to the two-photon subspace defined by ${\cal T}_p$, both $\hat{E}_0$ and $\hat{E}_1$ do not contain any basis states outside this subspace, such as $\ket{(HH)0}$. Furthermore, for input states restricted to the coincidence basis one can drop the last two terms from $\hat{E}_0$ and by monitoring the output of the FPBS in the coincidence basis only, allowing these states to be transmitted and rejecting all other cases, one can also drop the $\hat{E}_1$ operator to give the non-unitary non trace-preserving channel ${\cal E}_F (\rho)=\hat{E}_0 \rho \hat{E}_0^\dag$. This channel describes the combined action of the FPBS and monitoring in the coincidence basis with the Kraus operator
\begin{equation}
\hat{E}_0=\ketbra{HH}{HH}+\ketbra{VV}{VV}. \label{fusek}
\end{equation}
This is a parity check operation~\cite{Pitt,PAN} in the sense that only states where both of the photons have the same polarization (even parity) are allowed to be transmitted. This operation can be used as a fundamental component in a variety of quantum protocols, such as quantum error correction~\cite{NC}, entanglement purification~\cite{PAN} and filtering~\cite{Hof,OBRIEN}. It also forms part of an efficient method to generate cluster state resources that can be used to carry out quantum computation~\cite{RBH}. Indeed, the parity check is used in Type-I fusion for efficiently building one-dimensional~\cite{BR} and two-dimensional cluster states~\cite{BR,Yaakov} and can also be considered as a basic fusion operation; if successful, it matches Type-I fusion, but without the loss of any photons involved~\cite{PANPRA, BODDUAN}. In our experiment, however, in order to `monitor' the output photon number in the coincidence basis we detect the photons by destroying them. Thus our parity check, or fusion operation, is a postselected version and not scalable in its current form, unlike Type-I fusion~\cite{BR}. This is due to an exponential decrease in the total success probability when using many of these operations. It should be noted that active monitoring in the coincidence basis without destroying the photons can be achieved in principle using a non-demolition type measurement of the photon number~\cite{QND}, making this approach scalable using the techniques of Ref.~\cite{BR}. However, such measurements require the photons to pass through media with a large nonlinear response at the single-photon level, which is technically challenging at present and beyond the scope of this work~\cite{Guerlin}.

Ideally, the FPBS channel ${\cal E}_F$ from Eq.~(\ref{fusek}) takes two photons in the state $\rho_{in}=\ketbra{in}{in}$, with $\ket{in}=\ket{+}\ket{+}$ and $\ket{+}=\frac{1}{\sqrt{2}}(\ket{H}+\ket{V})$, and fuses them into the maximally entangled state $\ket{\phi^+}~=~(1/\sqrt{2})[\ket{HH}+\ket{VV}]$, equivalent to a two-qubit cluster state under a local Hadamard rotation on the first qubit. 
The above fusion operation occurs with success probability $p_0={\rm Tr}(\hat{E}_0 \rho_{in} \hat{E}_0^\dag)=1/2$ due to the non trace-preserving nature of the channel. Note that regardless of the success probability of a given input state $\rho_{in}$ being transmitted into the coincidence basis, the channel ${\cal E}_F$ always acts with unit probability. Either a coincidence is measured and the input state is transmitted into the coincidence basis, or no coincidence is measured and the input state is not transmitted into the coincidence basis. Moreover, if both photons represent the end qubits of two seperate linear cluster states, as shown in Fig.~\ref{setup}~{\bf (b)}, then after passing through the channel ${\cal E}_F$ the cluster states are fused together; here the two linear cluster states involved are joined with one of the photons representing the middle qubit between the edges bonding to the two clusters and the other photon representing a qubit directly attached to it~\cite{graph}. If failure occurs one tries again with the next qubits in both clusters~\cite{BR}. Two-dimensional cluster states can also be formed using the channel ${\cal E}_F$ as shown in Fig.~\ref{setup}~{\bf (c)}, or in a scalable way based on the techniques of Refs.~\cite{BR} and~\cite{Yaakov}.

\section{Fusion interference}
In order to check that our FPBS is implementing the correct fusion operation ${\cal E}_F$ and coherently interfering two input photons in the required spatio-temporal manner we set the input state of the two signal photons to $\ket{in}=\ket{+}\ket{+}$ and modify a time delay $\delta \tau$ between the input photon pulses using a translation stage, as shown in~Fig.~\ref{setup}. This allows us to maximize the coincidence probability temporally, with the FPBS single-mode input fibers ensuring good spatial overlap. At the detectors we measure the coincidence probability for the $\ket{+}\ket{+}$ state population, which for the expected entangled state $\ket{\phi^+}$ would give 1/2. However, the state $\ket{\phi^+}$ is produced from the fusion process acting on the input state $\ket{in}$ with a probability 1/2, so we expect the total coincidence probability to be 1/4. On the other hand, for an incoherent interference of the two photons,~{\it i.e.}~when $\delta \tau$ is larger than the mutual coherence time of the incoming photons $\tau_c$, but less than our detector coincidence time-window $\tau_{\rm coinc}$, the probability for a coincidence is 1/8 upon considering all possible outcomes of the non-interfering photons. Thus, as we modify the time delay $\delta \tau$ in our setup, we look for an `antidip' or peak in the coincidence probability (number of coincidences). This is a Hong-Ou-Mandel type interference effect~\cite{HOM} and a rigorous time-dependent derivation of the quantum interference phenomenon leading to this antidip in the coincidence probability is outlined in Appendix A. Here one finds that the probability for a coincidence at a set time delay of $\delta \tau$ is given by
\begin{equation}
P_{\rm coinc}(\delta \tau)=\frac{1}{8}(e^{- (\delta \tau / \sigma_t) ^2}+1) \label{coincp2},
\end{equation}
\begin{figure*}[t]
\begin{minipage}{15cm}
\hspace*{1cm}
\includegraphics[width=15cm]{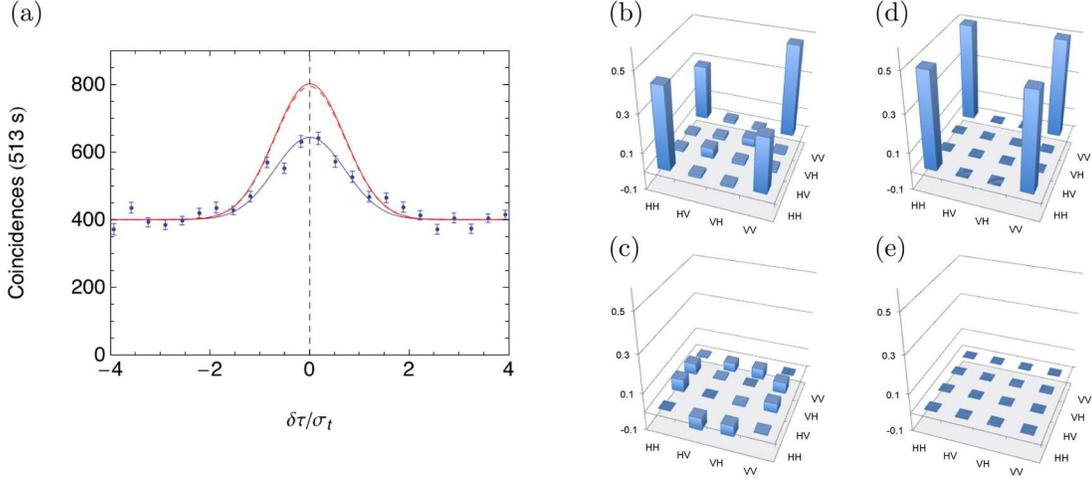}
\end{minipage}
\caption{Coincidences and tomographic reconstruction of the density matrix for the fused state resulting from input state $\ket{+}\ket{+}$. {\bf (a):}~Fourfold coincidences measured as a function of the delay time $\delta \tau$ (divided by the signal pulse duration $\sigma_t$) between the heralded signal photons clearly showing the expected antidip around zero delay. Using a relatively low pump power of 5.3 mW per fiber in this case, we obtain 23,000 pair coincidences/s per fiber and a maximum of 1.25 fourfold coincidences/s. The top curve is the ideal theoretical dependence and the dashed line includes a possible frequency mismatch corresponding to $\Delta \lambda_m =0.06$~nm for the signal photons. The lower solid line is a best fit of the data to the function $N_{\rm av}(p_0e^{- (\delta \tau / \sigma_t)^2}+1)/8$, finding $p_0=0.61$ and $N_{\rm av}$ is the average number of counts in our experiment when $\delta \tau \gg \sigma_t$. {\bf (b):}~Experimental real part of density matrix. {\bf (c):}~Experimental imaginary part. {\bf (d):}~Ideal real part for $\ket{\phi^+}$. {\bf (e):}~Ideal imaginary part for $\ket{\phi^+}$.}
\label{coinc}
\end{figure*}
where $\sigma_t$ is the pulse duration of the signal photon, defined as $2 \sqrt{2 \ln 2} / \Delta\omega$, and $\Delta\omega$ is the full width at half maximum of the signal's spectral intensity. In our experiment $\sigma_t\simeq 1$ps and $\Delta\omega$ corresponds to $\Delta\lambda\simeq0.5$nm. For complete temporal overlap $\delta \tau=0$ and for coherent interference we expect a coincidence probability of 1/4. On the other hand for large $\delta \tau$, $|\delta \tau| \gg \tau_c$, where $\tau_c \sim \sigma_t$, and with the coincidence time-window larger $\tau_{\rm coinc} \gg |\delta \tau|$ we have incoherent interference and the coincidence probability drops to 1/8 as mentioned above. 

In Fig.~\ref{coinc}~{\bf (a)} we show the expected number of coincidences $N_{\rm coinc}(\delta \tau)$ obtained from Eq.~(\ref{coincp2}). Here, $N_{\rm coinc}(\delta \tau)=N_{\rm av} P_{\rm coinc}(\delta \tau)$, where we have multiplied $P_{\rm coinc}(\delta \tau)$ by the average number of counts in our experiment when $|\delta \tau| \gg \sigma_t$ in order to scale it correctly. We have $N_{\rm av}=401$. The function $N_{\rm coinc}(\delta \tau)$ is plotted as the solid upper curve. In the lower dotted curve we include a possible frequency mismatch between the central frequencies of the two signal photons, $\omega^0_1$ and $\omega^0_2$, corresponding to $\Delta \lambda_m =0.06$~nm coming from the spectral resolution of our spectrometer. This is calculated by carrying out the necessary integration over the coincidence time-window for a spectral-temporal dependent version of Eq.~(\ref{coincp2}) derived in Appendix A as Eq.~(\ref{coincpdelta}). 

The data points from our experiment are also shown in Fig.~\ref{coinc}~{\bf (a)}, along with a best fit to the curve defined by $N_{\rm coinc}^{exp}(\delta \tau)=N_{\rm av} P^{fit}_{\rm coinc}(\delta \tau)$, where $P^{fit}_{\rm coinc}(\delta \tau)=(p_0e^{- (\delta \tau / \sigma_t)^2}+1)/8$, with $p_0=0.61$, giving an antidip visibility of 61\%. Here, the visibility is defined as the percentage ratio of $(N_{\rm max}-N_{\rm min})/N_{\rm min}$. In our experiment we have $\tau_{\rm coinc}=3$~ns, $\tau_{\rm rep}=12.5$~ns and $\delta \tau$ is varied using a translation stage on the pump beam before entering one of the PCFs, as shown in Fig.~\ref{setup}~{\bf (a)}. The errors are calculated using a Monte Carlo procedure with Poissonian fluctations in the count statistics~\cite{James}. The main causes of deviation from the ideal case are the spectral mixedness of the signal photons from our sources~\cite{Clark}, leading to broadening of the coincidence profile~\cite{LEG} and spatial mismatch between the interfering signal photons at the FPBS. We discuss later how these effects can be incorporated into our theoretical model. Note that photon loss in general affects the overall number of coincidences regardless of the time delay $\delta \tau$, leaving the visibility of the antidip unaffected.
\begin{table}
\caption{\label{tabone}Properties of the fused state as the pump power is increased.} 
\lineup
\begin{tabular}{@{}*{6}{l}}
\br                
            
\0\0{\footnotesize power (mW)}&{\footnotesize 4-fold rate ($s^{-1}$)}&{\footnotesize pairs/pulse}&\m \0\0{\footnotesize$F_{\phi^+}$}&\m \0\0 {\footnotesize $C$} & \0\0\0 {\footnotesize $P$} \cr 
\mr
\0\0\0\0\0 {\footnotesize 5.3}&\0\0\0\0 {\footnotesize 3.2} &\0\0 {\footnotesize 0.037}&{\footnotesize $0.740\pm 0.007$}&{\footnotesize $0.550\pm 0.014$} &{\footnotesize $0.63\pm 0.01$} \cr
\0\0\0\0\0 {\footnotesize 7.9}&\0\0\0\0 {\footnotesize 9.8} &\0\0 {\footnotesize 0.064}&{\footnotesize $0.677\pm 0.006$}&{\footnotesize $0.392\pm 0.012$}&{\footnotesize $0.520\pm 0.007$} \cr 
\0\0\0\0\0 {\footnotesize 10.5}&\0\0\0 {\footnotesize 36.4} &\0\0 {\footnotesize 0.103}&{\footnotesize $0.606\pm 0.007$}&{\footnotesize $0.265\pm 0.015$} &{\footnotesize $0.448\pm 0.008$} \cr 
\0\0\0\0\0 {\footnotesize 13.2}&\0\0\0 {\footnotesize 77.8} &\0\0 {\footnotesize 0.160}&{\footnotesize $0.554\pm 0.006$}&{\footnotesize $0.15\pm 0.01$} &{\footnotesize $0.392\pm 0.005$} \cr 
\0\0\0\0\0 {\footnotesize 14.8}&\0\0 {\footnotesize 111.6} &\0\0 {\footnotesize  0.193}&{\footnotesize $0.520\pm 0.004$}&{\footnotesize $0.07\pm 0.01$} &{\footnotesize $0.359\pm 0.003$} \cr 
\br
\end{tabular}
\end{table}

In Fig.~\ref{coinc}~{\bf (b)}~and~{\bf (c)} we show a tomographic reconstruction of the experimentally fused entangled state of the two signal photons at $\delta \tau =0$. This is obtained by measuring the qubits in all combinations of the bases $\{\ket{H/V},\ket{+/-},\ket{R/L} \}$~\cite{James}. Ideally this state is the maximally entangled state $\ket{\phi^+}$ shown in Fig.~\ref{coinc}~{\bf (d)}~and~{\bf (e)}. At $3.2$ four-fold coincidences per second our experimental state has a fidelity of $F_{\phi^+}=0.74 \pm 0.01$ with respect to $\ket{\phi^+}$, entanglement quantified by the concurrence of $C=0.55 \pm0.01$ and a purity of $P=0.63 \pm 0.01$. Some background can be seen to be present in the state from the non-zero $\ket{HV}$ and $\ket{VH}$ terms in Fig.~\ref{coinc}~{\bf (b)}, which we attribute to higher-order photon emission, other background processes such as Raman scattering, and imperfect polarization operations in our measurements. The background also gives rise to several non-zero terms in Fig.~\ref{coinc}~{\bf (c)}. In Appendix B we analyse the effects of higher-order photon emission from the four-wave mixing process and find that at this count rate we expect the fidelity to be limited to $F_{\phi^+}=0.80$. By increasing the pump power to $14.8$~mW we achieve a four-fold coincidence rate of $111.6$ per second with a fidelity of $F_{\phi^+}=0.520 \pm 0.004$, a non-zero concurrence of $C=0.07 \pm0.01$ and a purity of $P=0.359 \pm 0.003$. In Table {\ref{tabone}} we show how various parameters change as the pump power is increased and higher-order emission becomes increasingly significant. These generation rates and corresponding state properties show the potential of using PCFs for the controlled generation of larger multi-qubit entangled states for carrying out quantum protocols and investigating quantum phenomena with desired state qualities and success rates. 

Spatial-temporal mode mismatch, loss and higher-order photon emission are sources of error leading to deviation of our experimental data from the ideal case in both the coincidence probability and state tomography. While an explicit analysis of all factors leading to these effects is beyond the scope of the current work, we will show how it is possible to develop a theoretical model to describe the non-ideal fusion operation generating entangled states and connect it to the experimental antidip coincidence probability curve shown in Fig.~\ref{coinc}~{\bf (a)}.

\section{Fusion process tomography}

\begin{figure*}[t]
\begin{minipage}{13cm}
\hspace*{2.5cm}
\includegraphics[width=13cm]{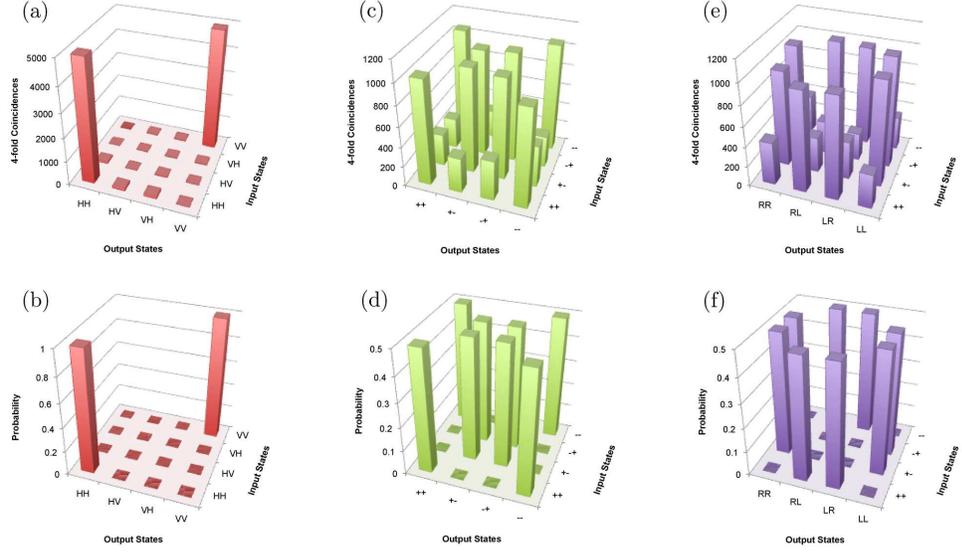}
\end{minipage}
\caption{Input and output states in the probing of the fusion operation. {\bf (a)}:  Experimental coincidences for the $Z \to Z$ basis. {\bf (b)}: Ideal probability for the $Z \to Z$ basis. The fidelity $F_{Z \to Z}$ is given by the ratio of correctly transmitted photon pairs to total number of transmitted pairs by the fusion operation for input basis $Z$ and output basis $Z$. Similarly for $F_{X \to X}$ and $F_{X \to Y}$. {\bf (c)}: Experimental coincidences for the $X \to X$ basis. {\bf (d)}: Ideal probability for the $X \to X$ basis. {\bf (e)}: Experimental coincidences for the $X \to Y$ basis. {\bf (f)}: Ideal probability for the $X \to Y$ basis. In all plots, each of the four-fold coincidence points was taken in $621$s.}
\label{fusion}
\end{figure*}

Ideally our fusion operation should act as the non trace-preserving channel ${\cal E}_F(\rho)=\hat{E}_0 \rho \hat{E}_0^\dag$, with $\hat{E}_0$ defined in Eq.~(\ref{fusek}). However, as the experimental fusion is not ideal, we must find a more approriate channel description. Assuming polarization changes from input to output states are negligible (see Fig.~\ref{fusion}~{\bf (a)}) we can write the experimental fusion operation as~\cite{OBRIEN}
\begin{equation}
{\cal E}_F^{exp}(\rho)= \sum_{n,m}\chi_{n,m}^F\hat{E}_n \rho \hat{E}_m^\dag \label{expchan}
\end{equation}
where the operators $\hat{E}_n$ are given as
\begin{eqnarray}
\hat{E}_0 & = & \ketbra{HH}{HH}+\ketbra{VV}{VV} \nonumber \\
\hat{E}_{zz} & = & \ketbra{HH}{HH}-\ketbra{VV}{VV} \nonumber \\
\hat{E}_{xy} & = & \ketbra{HV}{HV}+\ketbra{VH}{VH} \nonumber \\
\hat{E}_{xx} & = & \ketbra{HV}{HV}-\ketbra{VH}{VH}, 
\end{eqnarray}
which form an orthogonal operator set with respect to the Hilbert-Schmidt inner product. $\hat{E}_0$ is the ideal operation which occurs with probability $\chi_{0,0}^F$, the operator $\hat{E}_{zz}$ describes a phase flip error with probability $\chi_{zz,zz}^F$, the operator $\hat{E}_{xy}$ describes leakage of odd-parity states ($\ket{HV}$ and $\ket{VH}$) through the fusion device with probability $\chi_{xy,xy}^F$ and finally the operator $\hat{E}_{xx}$ describes leakage and a phase flip with probability $\chi_{xx,xx}^F$. Thus we have $n,m \in \{ 0,zz,xy,xx\}$ and $\sum_{n}\chi_{n,n}^F=1$. 
We can also define the {\it basis fidelity}~\cite{Hof2} for input basis states $i$ and output basis states $j$ as
\begin{equation}
F_{i \to j}= \frac{1}{2}\sum_{\ell,k} \bra{j_\ell}{\cal E}_F^{exp}(\ketbra{i_k}{i_k}) \ket{j_\ell}, \label{basfid}
\end{equation}
where for $F_{Z \to Z}$ we have $\{ \ket{HH},\ket{HV},\ket{VH},\ket{VV}\} \to \{ \ket{HH},\ket{VV}\}$, for $F_{X \to X}$ we have $\{ \ket{++},\ket{--}\} \to \{ \ket{++},\ket{--}\}$ and $\{ \ket{+-},\ket{-+}\} \to \{ \ket{+-},\ket{-+}\}$, for $F_{X \to Y}$ we have $\{ \ket{++},\ket{--}\} \to \{ \ket{LR},\ket{RL}\}$ and $\{ \ket{+-},\ket{-+}\} \to \{ \ket{LL},\ket{RR}\}$. Substituting the above $i$ and $j$ basis states into Eq.~(\ref{basfid}) and using the orthogonality of the $\hat{E}_n$ operators appearing in ${\cal E}_F^{exp}$ of Eq.~(\ref{expchan}), the basis fidelities are found to be equivalent to the elements of the $\chi$ matrix as follows~\cite{OBRIEN}
\begin{eqnarray}
F_{Z \to Z}&=& \chi_{0,0}^F+\chi_{zz,zz}^F \nonumber \\
F_{X \to X}&=& \chi_{0,0}^F+\chi_{xx,xx}^F \nonumber \\
F_{X \to Y}&=& \chi_{0,0}^F+\chi_{xy,xy}^F.
\end{eqnarray}
From these we then have the relation for the {\it process fidelity} of the fusion operation, $F_P^{exp} \equiv \chi_{0,0}^F$ describing the probability for successful fusion as
\begin{equation}
F_P^{exp}=\frac{1}{2}(F_{Z \to Z}+F_{X \to X}+F_{X \to Y}-1), \label{procfid3}
\end{equation}
where we have used $\sum_{n}\chi_{n,n}^F=1$. The basis fidelities can be found from our experiment using the relation
\begin{equation}
F_{i \to j}= \frac{\sum_{\ell,k} N^{out}_{i_k,j_\ell}}{\sum_{i_k}T_{i_k}}, \label{fidbas}
\end{equation}
as described in Appendix C. Here, $N^{out}_{i_k,j_\ell}$ is the number of times the output state is measured to be $\ket{j_\ell}$ when input state $\ket{i_k}$ is sent through device and $T_{i_k}=\sum_{i_p}N^{out}_{i_p,i_k}$. Eq.~(\ref{fidbas}) essentially means we can define the basis fidelities $F_{i \to j}$ to be the ratio of the total number of transmitted output states in the correct basis to the total number of transmitted states. In Fig.~\ref{fusion} we show the experimentally measured number of transmitted states (coincidences) for the input and output basis defined by the fidelities $F_{Z \to Z}$, $F_{X \to X}$ and $F_{X \to Y}$. We find $F_{Z \to Z}=0.958 \pm0.001$ from the data shown in Fig.~\ref{fusion}~{\bf (a)}, with the ideal transmission coincidence probabilities for $Z \to Z$ shown in Fig.~\ref{fusion}~{\bf (b)}. From the data corresponding to Fig.~\ref{fusion}~{\bf (c)} we find $F_{X \to X}=0.768 \pm 0.001$. The ideal transmission coincidence probabilities for $X \to X$ are shown in Fig.~\ref{fusion}~{\bf (d)}. Finally, from the data corresponding to Fig.~\ref{fusion}~{\bf (e)} we find $F_{X \to Y}=0.759 \pm 0.001$, with the ideal transmission coincidence probabilities for $X \to Y$ shown in Fig.~\ref{fusion}~{\bf (f)}. From these three fidelities we then find, using Eq.~(\ref{procfid3}), a process fidelity of our fusion operation of $F_P^{exp}=0.743 \pm 0.001$.

The entanglement capability of our fusion operation can be defined as the maximum entanglement that can be generated from it using input product states. We can quantify this in terms of the concurrence $C$ and we have the lower bound for the entanglement capability of the fusion operation, quantified by the concurrence as $C_E \ge 2 F_P^{exp}-1$, as shown in Appendix D. Based on the above process fidelity of our experiment we have $C_E \ge 0.485 \pm 0.002$. Note that both $F_P^{exp}$ and $C_E$ are consistent with the fidelity and concurrence calculated directly from the reconstructed density matrix of the last section for the rate $6.62$ events per second at $\delta \tau = 0$. A small deviation is due to the assumption of no polarization change from input to output states in the model used to describe the experimental fusion process ${\cal E}_F^{exp}$. 

\section{Non-ideal temporal setting}

So far we have considered our fusion operation at the maximum visibility $\delta \tau = 0$. In the ideal case of  $\delta \tau = 0$ we have that the state $\ket{+}_1\ket{+}_2$ is transformed into the entangled state $\ket{\phi^+}_{1'2'}$ with probability $1/2$. However, for $|\delta \tau| \gg \tau_c$ we obtain the incoherent mixture $\rho=\frac{1}{2}(\ketbra{HH}{HH}+\ketbra{VV}{VV})_{1'2'}$ with probability $1/2$. Here, the photon in mode $1'$ is detected before the photon in mode $2'$ (or vice-versa) and even though the time period $|\delta \tau|$ is outside the coherence time of the individual photons $ \tau_c$, the larger coincidence window $\tau_{\rm coinc}$ means we still register the state as being a valid output state from the fusion process. This loss of coherence in the output state suggests the process might be described by a phase damping channel acting on the output photons. This can be incorporated into the fusion channel by writing the total action of the fusion as ${\cal E}_T(\rho_{in})={\cal E}_{PD}({\cal E}_F(\rho_{in}))$, where ${\cal E}_{PD}$ is a phase damping channel, the form of which we derive next. The same channel might also describe other forms of decoherence, such as that resulting from imperfect spatial-mode overlap or spectral distinguishability between the interfering photons. These parameters are challenging to set in an experiment so it is useful to understand how the fusion operation performs under such conditions. By developing a model that describes the effect of non-ideal control we may then quantify the requirements for a given level of performance of the fusion. 

\begin{figure*}[t]
\begin{minipage}{14cm}
\hspace*{1.8cm}
\includegraphics[width=14cm]{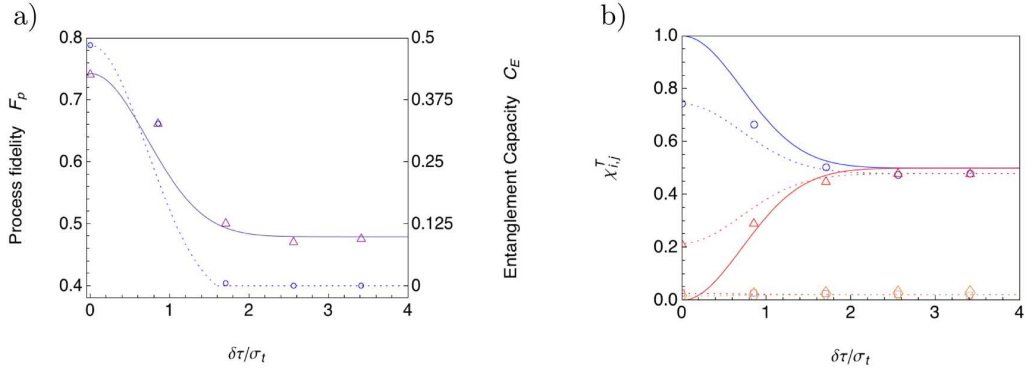}
\end{minipage}
\caption{The total process fidelity, entanglement capability and elements of the process matrix. {\bf (a):}~The total process fidelity $F_P^T$ and entanglement capability $C_E$ dependence on the time delay $\delta \tau$. The solid (dotted) line shows the decay expected from the theoretical phase damping model with the experimental fusion values taken at zero time delay for $F_P^T$ ($C_E$). The experimental values are plotted as triangles (circles), with errors negligible on the scale of the graph. {\bf (b):}~The elements of the total process matrix $\chi^T$ as they depend on the time delay $\delta \tau$. The upper (lower) solid line corresponds to the element $\chi_{0,0}^T$ ($\chi_{zz,zz}^T$) from the theoretical model with perfect fusion. The upper (lower) dashed line corresponds to the element $\chi_{0,0}^T$ ($\chi_{zz,zz}^T$) from the experimental data at $\delta \tau=0$ extrapolated using the function $f(\delta \tau)$. The lower lines show the elements $\chi_{xx,xx}^T$ and $\chi_{xy,xy}^T$. For ideal fusion these are zero regardless of the time delay. The experimental values are plotted as circles, triangles, squares and diamonds for $00$, $zz$, $xx$ and $yy$ respectively, with errors negligible on the scale of the graph. Note that all diagonal $\chi^T$ elements sum to one regardless of the time delay $\delta \tau$.}
\label{fidelity2}
\end{figure*}

For a single qubit, a phase damping channel with an arbitrary time dependent dephasing function $f(\delta \tau)\in [0,1]$ can be described by the channel ${\cal E}(\rho)=\sum_{i=0}^1 \hat{K}_i \rho \hat{K}_i^\dag$, with the Kraus operators 
\begin{eqnarray}
\hat{K}_0&=&\frac{1}{\sqrt{2}}[(1+f(\delta \tau))]^{1/2} \openone \nonumber \\
\hat{K}_1&=&\frac{1}{\sqrt{2}}[(1-f(\delta \tau))]^{1/2} \sigma_z, \nonumber
\end{eqnarray}
which satisfy $\hat{K}_0^\dag \hat{K}_0+ \hat{K}_1^\dag \hat{K}_1=\openone$. Usually the function $f(\delta \tau)=e^{-\Gamma \delta \tau}$ is considered, where $\Gamma$ represents some decay rate and $\delta \tau$ is a positive exposure time. Here we wish to determine the form of $f(\delta \tau)$ for our fusion operation initially using $\delta \tau$ as the temporal parameter. Considering two qubits subject to such a type of arbitrary phase damping, we have the overall channel ${\cal E}_{PD}(\rho)=\sum_{i=0}^3 \hat{K}_i \rho \hat{K}_i^\dag$, where 
\begin{eqnarray}
\hat{K}_0&=&\alpha^2 \openone \otimes \openone \nonumber \\
\hat{K}_1&=&\alpha \beta \openone \otimes \sigma_z \nonumber \\
\hat{K}_2&=&\alpha \beta \sigma_z \otimes \openone \nonumber \\
\hat{K}_3&=&\beta^2 \sigma_z \otimes \sigma_z,
\end{eqnarray}
with $\alpha=[(1+f(\delta \tau))/2]^{1/2}$ and $\beta=[(1-f(\delta \tau))/2]^{1/2}$. We can now write the total channel ${\cal E}_{T}(\rho)={\cal E}_{PD}({\cal E}_F(\rho))$ describing the non-ideal fusion and phase damping due to $\delta \tau \neq 0$ in the $\{ \hat{E}_n\}$ basis as
\begin{eqnarray}
{\cal E}_{T}(\rho)&=& \sum_{\ell,n,m}\chi_{n,m}^{F}\hat{K}_{\ell}\hat{E}_n \rho \hat{E}_m^\dag \hat{K}_{\ell}^\dag= \sum_{n,m}\chi_{n,m}^{T}\hat{E}_n \rho \hat{E}_m^\dag. \label{totalchan}
\end{eqnarray}
The second equality is possible because $\openone$ and $\sigma_z$ do not change the polarization, so that we can rewrite the combined operators more conveniently in terms of only $\hat{E}_{0}$, $\hat{E}_{zz}$, $\hat{E}_{xx}$, $\hat{E}_{xy}$, and the elements of a new matrix $\chi^{T}$ using the identities $\openone \otimes \openone \nonumber \equiv \hat{E}_0+\hat{E}_{xy}$, $\openone \otimes \sigma_z \nonumber \equiv \hat{E}_{zz}-\hat{E}_{xx}$, $\sigma_z \otimes \openone \nonumber \equiv \hat{E}_{zz}+\hat{E}_{xx}$ and $\sigma_z \otimes \sigma_z \nonumber \equiv \hat{E}_0-\hat{E}_{xy}$. In particular, we have that the total success probability for the fusion operation is given by $F_P^T=\chi_{0,0}^T=(\alpha^4 + \beta^4)\chi_{0,0}^{F}+(2\alpha^2\beta^2)\chi_{zz,zz}^{F}$. Substituting in for $\alpha$ and $\beta$, with $f(\delta \tau)$ arbitrary, we have
\begin{equation}
F_P^T=\frac{1}{2}(\chi_{0,0}^{F}+\chi_{zz,zz}^{F})+\frac{1}{2}f(\delta \tau)^2(\chi_{0,0}^{F}-\chi_{zz,zz}^{F}). \label{procfid}
\end{equation}
Similarly, we have
\begin{eqnarray}
\chi_{zz,zz}^T &=& \frac{1}{2}(\chi_{0,0}^{F}+\chi_{zz,zz}^{F})-\frac{1}{2}f(\delta \tau)^2(\chi_{0,0}^{F}-\chi_{zz,zz}^{F}) \nonumber \\
\chi_{xx,xx}^T &=&\frac{1}{2}(\chi_{xx,xx}^{F}+\chi_{xy,xy}^{F})+\frac{1}{2}f(\delta \tau)^2(\chi_{xx,xx}^{F}-\chi_{xy,xy}^{F}) \nonumber \\
\chi_{xy,xy}^T &=&\frac{1}{2}(\chi_{xx,xx}^{F}+\chi_{xy,xy}^{F})-\frac{1}{2}f(\delta \tau)^2(\chi_{xx,xx}^{F}-\chi_{xy,xy}^{F}). \nonumber \\
&& \label{procfid2}
\end{eqnarray}
We can find the expected form of the phase damping function $f(\delta \tau)$ by considering the coincidence probability $P_{\rm coinc}$ given in Eq.~(\ref{coincp2}) for the case of ideal fusion. Here, the fusion operation acts on the state $\ket{+}\ket{+}$ and produces the state
\begin{eqnarray}
\rho&=&\frac{1}{2}\big[\ketbra{HH}{HH}+\ketbra{HH}{VV}+\ketbra{VV}{HH}+\ketbra{VV}{VV}\big],
\label{fusestate}
\end{eqnarray}
with probability $1/2$. After phase damping we then end up with the state $\rho'={\cal E}_{PD}(\rho)$ given by
\begin{equation}
\hskip-1.3cm \rho'=\frac{1}{2}\big[\ketbra{HH}{HH}+f(\delta \tau)^2\ketbra{HH}{VV} +f(\delta \tau)^2\ketbra{VV}{HH}+\ketbra{VV}{VV}\big],
\end{equation}
Measuring the population of $\rho'$ in the state $\ket{+}\ket{+}$ gives $(1+f(\delta \tau)^2)/4$. By including the factor of $1/2$ due to the probabilistic nature of the fusion producing the state in Eq.~(\ref{fusestate}) we have that $P_{\rm coinc}=(1+f(\delta \tau)^2)/8$. Using Eq.~(\ref{coincp2}) we are able to then make the correspondence $f(\delta \tau)=e^{-(\delta \tau / \sigma_t)^2/2}$. In Fig.~\ref{fidelity2}~{\bf (a)} we show the {\it expected} dependence of the total process fidelity $F_P^T$ and entanglement capability with the time delay $\delta \tau$ in the case of our experimental fusion at $\delta \tau=0$. The data give a close match to the expected dependence. For definiteness we choose positive $\delta \tau$ with the temporal dependence being symmetric. In both of these we have used the phase damping function $f(\delta \tau)=e^{-(\delta \tau / \sigma_t)^2/2}$. 

In Fig.~\ref{fidelity2}~{\bf (b)} we show the $\chi^T$ matrix elements for the total process of fusion and phase damping, as defined in Eqs.~(\ref{procfid})~and~(\ref{procfid2}). The solid lines correspond to the expected dependence of the ideal fusion with phase damping and the dotted lines correspond to the expected dependence of our experimental fusion at $\delta \tau=0$ with phase damping applied. In both, the phase damping function is $f(\delta \tau)=e^{-(\delta \tau / \sigma_t)^2/2}$. Again, the data closely match the $f(\delta \tau)$ model. Note that for ideal fusion, the elements $\chi_{xx,xx}^T$ and $\chi_{xy,xy}^T$ are zero always, regardless of the amount of phase damping. However, Fig.~\ref{fidelity2}~{\bf (b)} shows that this is not the case for nonideal fusion, as in our experiment. Note also from Fig.~\ref{fidelity2}~{\bf (b)} that in the case of our experimental fusion at $\delta \tau = 0$, the effects of imperfections such as spectral mixedness and spatial mismatch can be almost entirely described in terms of an ideal fusion operation (solid lines) that has been phase damped. Indeed, at $\delta \tau/\sigma_t \simeq 1$ in Fig.~\ref{fidelity2}~{\bf (b)}, one can see that the phase-damped ideal fusion matches the non-damped experimental fusion at $\delta \tau =0$. Thus to a good approximation ($\chi_{xx,xx}^T$ and $\chi_{xy,xy}^T$ negligible), we can model our experimental fusion operation at $\delta \tau =0$ as being a phase damped ideal fusion operation. For $\delta \tau >0$ an additional phase damped channel with the function $f(\delta \tau)$ is then applied to each qubit.

\section{Summary}

In this work we reported an experimental demonstration of the fusion of photons from two independent photonic crystal fiber sources into polarization entangled states. We introduced and carried out a novel method to characterize the fusion operation via quantum process reconstruction. To do this we developed a theoretical model to describe the entangling process which included the main imperfections. We then showed that our model fitted well with the experimental results. This work highlights the need for accurate control of spatial and temporal properties for the success of photonic fusion and provides detailed information about the practical requirements and limitations of an operation that may be used in emerging quantum photonic technologies. Future work will be to optimize our fusion operation and use it to generate larger high-quality multi-qubit entangled states, including the use of additional degrees of freedom such as path and frequency encoding, for carrying out computations and communication protocols.

\ack
We thank Prof.~M.~S.~Kim for discussions and acknowledge support from UK EPSRC, EU project 248095 Q-Essence and ERC grant 247462 QUOWSS.

\appendix

\appendix
\section*{Appendix A: Derivation of fusion interference antidip}
\setcounter{section}{1}

Starting from the initial state $\ket{H}_1\ket{H}_{2}$ for the signal photons in modes 1 and 2, produced from the PCF sources as shown in~Fig.~\ref{setup}, and finishing with the measurement of the coincidence probability for the state $\ket{H}_{1a}\ket{H}_{2a}$ at the detectors $D1a$ and $D2a$, we have the probability density for a photon to be detected in mode $1a$ at time $t_0$ and another in mode $2a$ at time $t_0+\tau$ given by
\begin{equation}
\hskip -1 cm P_{\rm coinc}(t_0,\tau)= \tensor*[_{12}]{\bra{HH}}{} \hskip -0.1 cm\hat{E}^-_{H_{1a}}(t_0)\hat{E}^-_{H_{2a}}(t_0 \hskip -0.05 cm + \hskip -0.05 cm \tau)\hat{E}^+_{H_{2a}}(t_0 \hskip -0.05 cm + \hskip -0.05 cm \tau)\hat{E}^+_{H_{1a}}(t_0) \hskip -0.05 cm \ket{HH}_{12} 
\label{coincp}
\end{equation}
where the scalar photon-unit time-dependent electric field operators are defined as 
\begin{eqnarray}
\hat{E}^+_{i}(t)&=& \frac{1}{\sqrt{2 \pi}}\int {\rm d}\omega e^{-i \omega t} \hat{a}_i(\omega) \nonumber \\
\hat{E}^-_{i}(t)&=& \frac{1}{\sqrt{2 \pi}}\int {\rm d}\omega e^{i \omega t} \hat{a}_i^\dag(\omega) \label{expec}
\end{eqnarray}
with $\hat{a}_i^\dag(\omega)$ and $\hat{a}_i(\omega)$  the photon creation and annihilation operators respectively, which obey the bosonic commutation relation $[\hat{a}_i(\omega),\hat{a}_{j}^\dag(\omega')]=\delta_{ij}\delta(\omega-\omega')$. The initial state $\ket{HH}_{12}$ is written as
\begin{equation}
\ket{HH}_{12}=\int {\rm d}\omega_1 \hskip -0.1 cm \int {\rm d}\omega_2 \hskip 0.1 cm \phi_1(\omega_1) \phi_2(\omega_2) \hat{a}_{H_1}^\dag(\omega_1) \hat{a}_{H_2}^\dag(\omega_2) \ket{0}, \label{input}
\end{equation}
where $\phi_i(\omega_i)$ is the spectral amplitude of a single photon pulse in mode $i$, normalized so that $\int{\rm d}\omega \phi_i^*(\omega_i)\phi_i(\omega_i)=1$.
The initial state $\ket{H}_1\ket{H}_{2}$ is rotated into the required input state $\ket{in}=\ket{+}_1\ket{+}_2$ of the FPBS by HWPs that produce the transformations $\hat{a}_{H_1}^\dag(\omega_1) \to (1/\sqrt{2})[\hat{a}_{H_1}^\dag(\omega_1)+\hat{a}_{V_1}^\dag(\omega_1)]$ and $\hat{a}_{H_2}^\dag(\omega_2) \to (1/\sqrt{2})[\hat{a}_{H_2}^\dag(\omega_2)+\hat{a}_{V_2}^\dag(\omega_2)]$. The action of the FPBS, combined with the HWP-QWP-HWP chain on mode 2 (for phase correction), produces the transformations 
\begin{eqnarray}
\hat{a}_{H_1}^\dag(\omega_1)& \to & ~~\hat{a}_{H_{2'}}^\dag(\omega_1) \label{pbstrans} \\
\hat{a}_{V_1}^\dag(\omega_1)& \to & ~~i \hat{a}_{V_{1'}}^\dag(\omega_1) \nonumber \\
\hat{a}_{H_2}^\dag(\omega_2)& \to & ~~\hat{a}_{H_{1'}}^\dag(\omega_2) \nonumber \\
\hat{a}_{V_2}^\dag(\omega_2)& \to & -i \hat{a}_{V_{2'}}^\dag(\omega_2). \nonumber 
\end{eqnarray}
Next, the HWP and QWP on each mode produce the transformations
\begin{eqnarray}
\hat{a}_{H_{1'}}^\dag(\omega_i)& \to &(1/\sqrt{2})[\hat{a}_{H_{1'}}^\dag(\omega_i)+\hat{a}_{V_{1'}}^\dag(\omega_i)] \nonumber \\
\hat{a}_{V_{1'}}^\dag(\omega_i)& \to &(1/\sqrt{2})[\hat{a}_{H_{1'}}^\dag(\omega_i)-\hat{a}_{V_{1'}}^\dag(\omega_i)] \nonumber \\
\hat{a}_{H_{2'}}^\dag(\omega_i)& \to &(1/\sqrt{2})[\hat{a}_{H_{2'}}^\dag(\omega_i)+\hat{a}_{V_{2'}}^\dag(\omega_i)] \nonumber \\
\hat{a}_{V_{2'}}^\dag(\omega_i)& \to &(1/\sqrt{2})[\hat{a}_{H_{2'}}^\dag(\omega_i)-\hat{a}_{V_{2'}}^\dag(\omega_i)]. \nonumber  
\end{eqnarray}
The PBSs that follow produce the transformations
\begin{eqnarray}
\hat{a}_{H_{1'}}^\dag(\omega_i)& \to & ~~\hat{a}_{H_{1a}}^\dag(\omega_i) \nonumber \\ 
\hat{a}_{V_{1'}}^\dag(\omega_i)& \to & ~~i \hat{a}_{V_{1b}}^\dag(\omega_i) \nonumber \\
\hat{a}_{H_{2'}}^\dag(\omega_i)& \to & ~~\hat{a}_{H_{2a}}^\dag(\omega_i) \nonumber \\
\hat{a}_{V_{2'}}^\dag(\omega_i)& \to & -i \hat{a}_{V_{2b}}^\dag(\omega_i). \nonumber 
\end{eqnarray}
Thus, only horizontally polarized photons will be detected at detectors $D1a$ and $D2a$. These last operations (the HWP, QWP and PBS) together with the detections in modes $1a$ and $2a$ are equivalent to measuring the coincidence probability for the state $\ket{+}_{1'}\ket{+}_{2'}$ in the output of the fusion. Therefore $P_{\rm coinc}(t_0,\tau)$ in Eq.~(\ref{coincp}) represents the coincidence probability density for the state $\ket{+}_{1'}\ket{+}_{2'}$.

Substituting all the above transforms into Eq.~(\ref{input}), then substituting this into Eq.~(\ref{coincp}) with the definitions given in Eq.~(\ref{expec}) and carrying out the necessary integrations~\cite{LEG} one finds 
\begin{equation}
P_{\rm coinc}(t_0,\tau)=\frac{1}{16}| \zeta_1(t_0+\tau) \zeta_2(t_0)+\zeta_1(t_0)\zeta_2(t_0+\tau)|^2, \label{coincprob}
\end{equation}
where $\zeta_i(t)=\frac{1}{\sqrt{2 \pi}}\int {\rm d}\omega_i \phi_i(\omega_i)e^{-i \omega_i t}$ is the spatio-temporal modefunction for mode $i$. Choosing Gaussian single-photon pulses for each of the input modes, with a time delay of $\delta \tau$ between the peaks of the pulses, we have
\begin{eqnarray}
\zeta_1(t)&=&(2/\pi)^{1/4} e^{-(t-\delta \tau/2)^2-i \omega^0_1t} \\
\zeta_2(t)&=&(2/\pi)^{1/4} e^{-(t+\delta \tau/2)^2-i \omega^0_2t}.
\end{eqnarray}
Here, $\omega^0_i$ is the central carrier frequency for mode $i$ expressed in units of $1/\sigma_t$ and the times $t$ and $\delta \tau$ are expressed in units of $\sigma_t$ ($\sigma_t$ is the pulse duration of the signal photon, defined as $2 \sqrt{2 \ln 2} / \Delta\omega$, and $\Delta\omega$ is the full width at half maximum of the signal's spectral intensity). Substituting these expressions into Eq.~(\ref{coincprob}) and integrating over all possible detection time $t_0$,~{\it i.e.}~$\int_{- \infty}^\infty {\rm d} t_0$, one finds the probability density for a coincidence at a time duration $\tau$ with a set delay of $\delta \tau$ given by
\begin{equation}
P_{\rm coinc}(\tau,\delta \tau)=\frac{e^{-\delta \tau^2-\tau^2}}{\sqrt{64 \pi}}(\cos[\tau(\omega^0_1-\omega^0_2)]+\cosh[2 \delta \tau\,\tau]). \label{coincpdelta}
\end{equation}
Setting the central carrier frequencies the same for the moment, $\omega^0_1=\omega^0_2$, we then integrate Eq.~(\ref{coincpdelta}) with respect to $\tau$ over the coincidence window. Here, the integration is $\int_{-\tau_{\rm coinc}/2}^{\tau_{\rm coinc}/2}{\rm d} \tau$, allowing for either signal photon to be detected first. For $\tau_{rep} > \tau_{\rm coinc} \gg |\delta \tau|$, where $\tau_{rep}$ is the time between the pump pulses (and therefore possible signal photons produced from four-wave mixing in the PCFs), one finds the coincidence probability
\begin{equation}
P_{\rm coinc}(\delta \tau)=\frac{1}{8}(e^{- (\delta \tau / \sigma_t)^2}+1) \label{acoincp2}.
\end{equation}
Where we have now included $\sigma_t$ explicitly. Although we have neglected a time-dependent evaluation of the idler photons in the above analysis, they can also be included. Indeed, it is straightforward to check that such a calculation does not change the result of Eq.~(\ref{acoincp2}), given that the idler photons are assumed to be in a product state with the signal photons. 

\appendix
\section*{Appendix B: Higher-order emission analysis}
\setcounter{section}{2}

The state generated by four-wave mixing in one source can be written as~\cite{Fulc}
\begin{equation}
\ket{\psi}={\cal N}(\ket{0,0}_{s,i}+\alpha\ket{H,H}_{s,i} + \alpha^2\ket{2H,2H}_{s,i}+O(\alpha^3)),
\end{equation}
where ${\cal N}$ is a normalisation constant and $|\alpha|^2 = \overline{n}$ is the mean number of signal-idler pairs generated in a pulse. When the idler detector registers a click, the heralded density matrix of the signal mode becomes 
\begin{equation}
\rho_s={\rm Tr}_i\left(\frac{\Pi_{\rm click}^{1/2}\ketbra{\psi}{\psi}\Pi_{\rm click}^{1/2}}{{\rm Tr}(\Pi_{\rm click}\ketbra{\psi}{\psi})}\right),
\label{signalden}
\end{equation}
where the action of the detector is described by the positive operator-valued measure $\{ \Pi_{\rm click}, \Pi_{\rm no-click}=\openone-\Pi_{\rm click}\}$, with $\Pi_{\rm click}=\sum_{n=0}^{\infty}(1-(1-\eta)^n)\ket{nH}_i\bra{nH}$ and $\eta$ is the lumped detector efficiency of registering a click given a single photon is input into the mode that it monitors. From Eq.~(\ref{signalden}) we have
\begin{equation}
\rho_s={\cal N}(\eta_1 \ket{H}\bra{H} + \overline{n}\eta_2\ket{2H}\bra{2H}+O(\overline{n}^2)),
\end{equation} 
where $\eta_n=1-(1-\eta)^n$. We now write the heralded state of the two sources to first-order in $\overline{n}$ as
\begin{equation}
\hskip-2cm \rho_{12}={\cal N}(\ket{H,H}_{12}\bra{H,H} + 2\overline{n}\gamma\ket{H,2H}_{12}\bra{H,2H}+2\overline{n}\gamma\ket{2H,H}_{12}\bra{2H,H})
\label{higherorderdensitymatrix},
\end{equation}
where $\gamma=\eta_2/2\eta_1$. The signal photons from $\rho_{12}$ are first rotated from horizontal to diagonal polarization then input to the fiber polarizing beamsplitter (FPBS). Using the transformations given in Eq.~(\ref{pbstrans}) of Appendix A we can write the output from the FPBS by making the following substitution in Eq. (\ref{higherorderdensitymatrix}) 
\begin{equation}
\ket{nH, mH}_{12}\to \frac{(\hat{a}^\dag_{H_{2'}}+i\hat{a}^\dag_{V_{1'}})^n(\hat{a}^\dag_{H_{1'}}-i\hat{a}^\dag_{V_{2'}})^m }{2^{(n+m)/2}\sqrt{n!m!}}.
\end{equation}
Where we have now switched from the Schr\"odinger to the Heisenberg picture. By removing the terms which do not lead to a possible coincidence of detection clicks between the two signal modes, then performing the polarization rotations used in analysing each measurement basis $\{ \ket{H/V},\ket{+/-}, \ket{R/L}\}$ and switching back into the Schr\"odinger picture to determine the contribution to the coincidence counts of the $O(\overline{n})$ terms in Eq.~(\ref{higherorderdensitymatrix}), we can evaluate the effects of higher-order emissions in our data. For instance, we find the visibility of the anti-dip in Fig.~\ref{coinc}, up to $O(\overline{n})$, is limited to
\begin{equation}
p_0=\frac{1-4\overline{n}\gamma+6\overline{n}\gamma^2}{1+6\overline{n}\gamma+3\overline{n}\gamma^2}.
\end{equation}
In addition, for the lowest power in Table 1 where $\overline{n}=0.037$, taking $\eta=0.1$ and considering up to $O(\overline{n})$ terms, the expected fidelity of the output state with respect to $\ket{\phi^+}$ is $F_{\phi^+}=0.80$. This assumes the fusion process is ideal, and so represents an upper bound.

\appendix
\section*{Appendix C: Derivation of basis fidelities}
\setcounter{section}{3}

The basis fidelities can be found from our experiment by noting the relation
\begin{equation}
\bra{j_\ell}{\cal E}_F^{exp}(\ketbra{i_k}{i_k}) \ket{j_\ell}=\frac{N^{out}_{i_k,j_\ell}}{N^{in}_{i_k}},
\end{equation}
where $N^{in}_{i_k}$ is the number of times input state $\ket{i_k}$ is sent through the device and $N^{out}_{i_k,j_\ell}$ is the number of times the output state is measured to be $\ket{j_\ell}$ when input state $\ket{i_k}$ is sent through device. We then have that
\begin{equation}
F_{i \to j}=\frac{1}{2}\sum_{\ell,k} \frac{N^{out}_{i_k,j_\ell}}{N^{in}_{i_k}}.
\end{equation}
Keeping the value of $N^{in}_{i_k}$ constant over the experiment for all possible input states $\ket{i_k}$,~{\it i.e.}~$N^{in}_{i_k}=N^{in}$, and noting that ${\rm Tr}({\cal E}_F^{exp}(\openone/4))=1/2$ for any experimental process matrix $\chi$ appearing in Eq.~(\ref{expchan}), we have that $N^{in}=\sum_{i_k}T_{i_k}/2$, where $T_{i_k}=\sum_{i_p}N^{out}_{i_p,i_k}$. In other words, for a given input basis $i$, the total number of transmitted states over that basis is equal to half the total number of input states (in this case $2 N^{in}$). Therefore we have
\begin{equation}
F_{i \to j}= \frac{\sum_{\ell,k} N^{out}_{i_k,j_\ell}}{\sum_{i_k}T_{i_k}}.
\end{equation}

\appendix
\section*{Appendix D: Relation of concurrence to fusion operation fidelity}
\setcounter{section}{4}

The concurrence, C, for a two qubit state $\rho_{12}$ can be written as~\cite{vert}
\begin{equation}
C(\rho_{12})={\rm max} \{ 0,  -\hskip -0.2cm \min_{A \in {\rm Sl}(2,{\mathbb C})}\hskip -0.2cm {\rm Tr}((\ketbra{A}{A})^{T_2} \rho_{12})\}, \label{conc1}
\end{equation}
where $\ket{A}$ denotes the unnormalized state $(A\otimes \openone)\ket{I}$, with $\ket{I}=\sum_i\ket{ii}$ and $A$ is any matrix with ${\rm det}(A)=1$. In Eq.~(\ref{conc1}) any choice for the state $\ket{A}$ provides a lower bound on the concurrence, thus choosing $\ket{A}=\sqrt{2}\ket{\psi^-}$, we have that $(\ketbra{A}{A})^{T_2}=\openone - 2 \ketbra{\phi^+}{\phi^+}$, and we can rewrite Eq.~(\ref{conc1}) as
\begin{equation}
C(\rho_{12}) \ge \max \{ 0, 2 F_{\phi^+}-1 \},
\end{equation}
where $F_{\phi^+}={\rm Tr}(\ketbra{\phi^+}{\phi^+} \rho_{12})$ is the fidelity of the state $\rho_{12}$ with respect to $\ket{\phi^+}$. If we consider the state $\rho_{12}$ to be an output state from a channel ${\cal E}$ acting on two input product states, then $C(\rho_{12})$ provides a lower bound on the entanglement capacity of that channel. 
Choosing the input product state to be $\rho_{in}=\ketbra{+}{+} \otimes \ketbra{+}{+}$ and the channel to be our experimental fusion operation ${\cal E}_F^{exp}$, we have that $\rho_{12}={\cal E}_F^{exp}(\rho_{in})=F_P^{exp}\ketbra{\phi^+}{\phi^+}+(1-F_P^{exp})\rho'$. Thus, $F_{\phi^+}=F_P^{exp}$ and we have the lower bound for the entanglement capability of the fusion operation, quantified by the concurrence as $C_E \ge 2 F_P^{exp}-1$.

\end{document}